\documentclass{article}

    \PassOptionsToPackage{numbers, compress}{natbib}

\usepackage{neurips_data_2024}

\usepackage[utf8]{inputenc} 
\usepackage[T1]{fontenc}    
\usepackage{hyperref}       
\usepackage{url}            
\usepackage{booktabs}       
\usepackage{amsfonts}       
\usepackage{nicefrac}       
\usepackage{microtype}      
\usepackage{xcolor}         
\usepackage[normalem]{ulem}
\usepackage{soul}
\usepackage{graphicx}
\usepackage{enumitem}
\usepackage{multirow} 
\usepackage{adjustbox}
\usepackage{amsmath}
\usepackage{tcolorbox}

\newcommand{\sectioncolor}{violet}

\title{A Benchmark Dataset for Multimodal Prediction of Enzymatic Function Coupling DNA Sequences and Natural Language}

\author{%
  Yuchen Zhang$^{1,3}$ \quad Ratish Kumar Chandrakant Jha$^{1*}$ \quad Soumya Bharadwaj$^{1*}$ \\\quad  \textbf{Vatsal Sanjaykumar Thakkar}$^{1*}$ \quad \textbf{Adrienne Hoarfrost}$^{2,4}$ \quad \textbf{Jin Sun}$^{1,4}$ \\
  $^1$School of Computing \quad 
  $^2$Department of Marine Sciences \\
  $^3$Department of Epidemiology \& Biostatistics \quad
  $^4$Institute for Artificial Intelligence\\
  University of Georgia \\
  \text{*} Equal contribution.
}

\begin{document}

\maketitle

\nolinenumbers

\begin{abstract}
Predicting gene function from its DNA sequence is a fundamental challenge in biology. Many deep learning models have been proposed to embed DNA sequences and predict their enzymatic function, leveraging information in public databases linking DNA sequences to an enzymatic function label. However, much of the scientific community's knowledge of biological function is not represented in these categorical labels, and is instead captured in unstructured text descriptions of mechanisms, reactions, and enzyme behavior. These descriptions are often captured alongside DNA sequences in biological databases, albeit in an unstructured manner. Deep learning of models predicting enzymatic function are likely to benefit from incorporating this multi-modal data encoding scientific knowledge of biological function. There is, however,  no dataset designed for machine learning algorithms to leverage this multi-modal information. Here we propose a novel dataset and benchmark suite that enables the exploration and development of large multi-modal neural network models on gene DNA sequences and natural language descriptions of gene function. We present baseline performance on benchmarks for both unsupervised and supervised tasks that demonstrate the difficulty of this modeling objective, while demonstrating the potential benefit of incorporating multi-modal data types in function prediction compared to DNA sequences alone. Our dataset is at \url{https://hoarfrost-lab.github.io/BioTalk/}.
\end{abstract}

\section{Introduction and motivation}\label{sec:Introduction}

    Identifying enzymatic function of gene sequences is a central task in biology. AI-driven methods for predicting function from sequence rely heavily on reference databases, which contain DNA sequences associated with functional annotation labels. These annotation labels represent scientific knowledge about a particular enzymatic or functional process, which are often summarized in these biological databases as unstructured text. Complicating this modeling task is the fact that biological databases contain known bias toward well-studied organisms, do not capture the full functional diversity of the natural world, and are exceptionally imbalanced with respect to annotation labels. An effective AI-driven functional prediction model therefore must be able to generalize to out of distribution sequences to reason about the potential function of novel DNA sequences. 

    Incorporating natural language descriptions of scientific knowledge about the function of a particular enzyme class alongside DNA sequences in multimodal frameworks holds promise for enhancing the embedding quality of DNA sequences and improving function prediction, particularly in out-of-distribution settings. Natural language descriptions provide rich, unstructured insights into enzyme mechanisms, reactions, and behavior, which, when integrated with DNA sequence data, are likely to enhance the predictive power of deep learning models. This approach also enables models to describe predicted functions in text, increasing interpretability and accessibility to biological researchers. 

    The development of better algorithms for multimodal prediction extends benefits beyond the field of biology to the broader machine learning community. Multimodal learning frameworks
    offer new insights into handling heterogeneous data and improving model generalization. By addressing the unique challenges of biological data, such as its heterogeneity and the need for high-quality curation, this research can inform advancements in other domains where multimodal data integration is critical.
    
    In recent years, the integration of multimodal data has significantly advanced various fields in machine learning \cite{gpt2,gpt3,llama3modelcard}. Despite the progress in this field, there is a lack of integrated AI-ready datasets that combine DNA sequences with their functional descriptions, limiting the development of advanced multimodal models coupling biological sequences and scientific knowledge captured in natural language. To address these critical needs, we introduce a novel multimodal dataset that combines DNA sequences of genes with text descriptions of their molecular function. This dataset is designed to enable the development of sophisticated multimodal models capable of predicting the functions of DNA sequences and providing detailed textual explanations.
    
    \noindent\textbf{Key contributions:}\label{subsec:Intro-KeyContri}
        \begin{enumerate}[leftmargin=*]
            \item \textbf{Novel Dataset}: We present a unique and comprehensive dataset that pairs DNA sequences with their corresponding functional descriptions, filling a critical gap in existing resources.
            \item \textbf{Multimodal Applications}: Our dataset facilitates the development of multimodal language models that can predict the functions of DNA sequences in detailed natural language descriptions.
            \item \textbf{Unimodal and Multimodal Benchmarks}: In addition to supporting multimodal applications, our dataset offers benchmarks for unimodal and multimodal models, including pretraining encoder-only transformer models on DNA sequences to enhance their performance on various tasks.
            \item \textbf{Impact}: The dataset facilitates the creation of DNA-language models for expansive applications, including functional prediction, sequence `captioning', and natural language design of novel genes. These capabilities would significantly enhance the interpretability and utility of genomic data. 
        \end{enumerate}

    \noindent\textbf{Open Access and Availability:}\label{subsec:Intro-website}
    All datasets and code are freely available. Details on access can be found at the accompanying website: \url{https://hoarfrost-lab.github.io/BioTalk/}.
\section{Background and related work}
    \noindent\textbf{Biological database content and structure}. 
        Biological "omics" data has accumulated at a vast scale since next generation sequencing technologies became widely available. The sequence data itself is highly structured and organized, linking associated proteins, genes, and genomes or metagenomes with their respective sequences. While highly structured, there is a known bias toward model organisms and biomedical annotations, resulting in databases which are not representative of global biodiversity and which are extremely imbalanced in the functions they represent \cite{lloyd}. These sequences are nonetheless annotated with information about the enzymatic function where appropriate, and associated behavior or reactions catalyzed by enzymes of that class, associated in databases describing Enzyme Commission (EC) numbers, which are hierarchical labels that categorize enzyme functions  \cite{explorenz}. However, this valuable information is often described in unstructured natural language text. 
        
        The Universal Protein Resource (UniProt) \cite{uniprot2021} and the European Nucleotide Archive (ENA) \cite{ENA} are major databases for storing and organizing protein and nucleotide sequences, respectively. UniProt is divided into UniProtKB/TrEMBL, which contains unreviewed protein sequences including those with computationally generated annotations, and UniProtKB/Swiss-Prot, which consists of manually curated records with empirical evidence of  functional information extracted from the literature and curator-evaluated computational analysis \cite{uniprot2021}. As with many biological databases with computationally inferred annotations, the UniProt database contains many more entries than SwissProt but also contains many functional misannotations, while the SwissProt database represents a smaller "gold standard" dataset \cite{misannotation}. The ENA database links gene DNA sequences to their corresponding UniProt protein IDs \cite{ENA}, and the Kyoto Encyclopedia of Genes and Genomes (KEGG) database describes high-level functions and utilities of biological systems by mapping DNA and protein sequences to biological pathways and enzyme functions within these pathways \cite{KEGG,kanehisa2023kegg,kanehisa2019toward}.        
    
    \noindent\textbf{Large DNA embedding models and functional prediction}.
    Most biological information, particularly that representing the global biodiversity that is poorly represented in reference databases, is available in the form of DNA sequences as raw reads or assembled genes or genomes. While much attention has been given to encoding protein sequences \cite{esm2, alphafold}, predicting function from gene sequence is perhaps a more central task to predicting novel biological functions.  We describe here two key published gene embedding models as well as our own DNA encoding model trained \textit{de novo} for comparison against our benchmarks.  
    
    \noindent\textit{\textbf{LOLBERT}}. We provide a model trained \textit{de novo} which we call Language of Life BERT (LOLBERT). LOLBERT is a BERT-based model pre-trained on a corpus of bacterial and archaeal genomic data from the Genome Taxonomy Database (GTDB) (release R214.0)~\cite{gtdb}. It utilizes a bidirectional transformer architecture with 12 self-attention layers and 12 hidden layers in its encoder. 
    
    \noindent\textit{\textbf{FinetunedLOLBERT}}. We further fine tuned LOLBERT specifically for the
    enzyme function classification task
    on the benchmarks
    (see Section~\ref{subsec:benchmark}). 

    \noindent\textit{\textbf{DNABERT}}. is a pre-trained LLM specifically designed for analyzing DNA sequences \cite{ji2021dnabert}.  It employs a masked language modeling approach similar to BERT, and is trained on a large collection of human genome sequences to predict masked nucleotides within the sequence \cite{ji2021dnabert}. This approach allows DNABERT to capture contextual relationships between nucleotides and understand the overall function of DNA sequences.

    \noindent\textit{\textbf{Nucleotide transformer}}. is introduced in \cite{dalla2023nucleotide},  another transformer based foundation model trained on DNA sequences from the human reference genome and genomes from several model organisms.

    \noindent\textbf{Multi-modal learning}. GPT-family models \cite{gpt2,gpt3} have demonstrated impressive performance in language tasks. Inspired by their success, multi-modal learning \cite{akkus2023multimodal} models have been proposed to leverage the understanding and reasoning ability of language models \cite{llama3modelcard} across different domains such as image \cite{liu2023llava}, audio \cite{akbari2021vatt}, autonomous vehicles \cite{cui2024survey}, and healthcare \cite{healthcare}. Importantly, learning on heterogeneous data allows a model to learn effective feature representations jointly \cite{clip} across domains that can be used for multiple downstream tasks. In a similar manner, we expect our benchmark datasets to enable the development of multi-modal models with improved capabilities for representing, predicting, and interpreting biological information.
\section{Benchmark datasets}\label{sec:dataset-n-benchmark}

    In this section, we detail data sources and splitting strategy for generating benchmark datasets of paired gene DNA sequences and associated natural language functional descriptions for training and evaluation of multi-modal DNA-language models.  
    
    \subsection{Dataset}\label{subsec:dataset}
    
        Our data are derived from the Universal Protein Resource (UniProt) \cite{uniprot2021,wang2021crowdsourcing} and their associated gene DNA sequences in the European Nucleotide Archive (ENA) \cite{ENA}. The UniProt database is bifurcated into two main sections: Uniprot/TrEMBL has a vast amount of data and greater functional diversity, but has more errors due to homology-based annotations whereas Uniprot/Swiss-Prot is a manually curated database with experimental evidence of all functional annotations.
        
        \noindent\textbf{Organism Filtering:} 
            Prokaryotes, comprising Bacteria and Archaea, are among the most abundant and diverse organisms on Earth. For this study, we filtered the data retrieved from UniProt to include only prokaryotic organisms. This focus allows us to explore a wide range of biological processes and enzymatic functions dominant in these domains.
        
        \noindent\textbf{Data Mapping, Retrieval, and Cleaning:} 
            Using ID mapping files\cite{uniprot2021}\cite{wang2021crowdsourcing} from UniProtKB, we mapped UniProt accession numbers to corresponding identifiers in UniRef (Universal Protein Resource Reference Clusters) UniRef50, UniRef90, and UniRef100, which define protein clusters based on amino acid sequence similarity at 50, 90, and 100 percent amino acid identity respectively; and to EMBL CDS IDs corresponding to gene coding sequences in the ENA database. The DNA sequences associated with each EMBL CDS ID were subsequently retrieved from ENA. 
            Records with missing or partial EC numbers were removed. Records containing multiple EC numbers or EMBL CDS IDs were separated into individual entries. 
            
            \begin{figure}[t]
                \flushleft(a)

                \includegraphics[width=\linewidth]{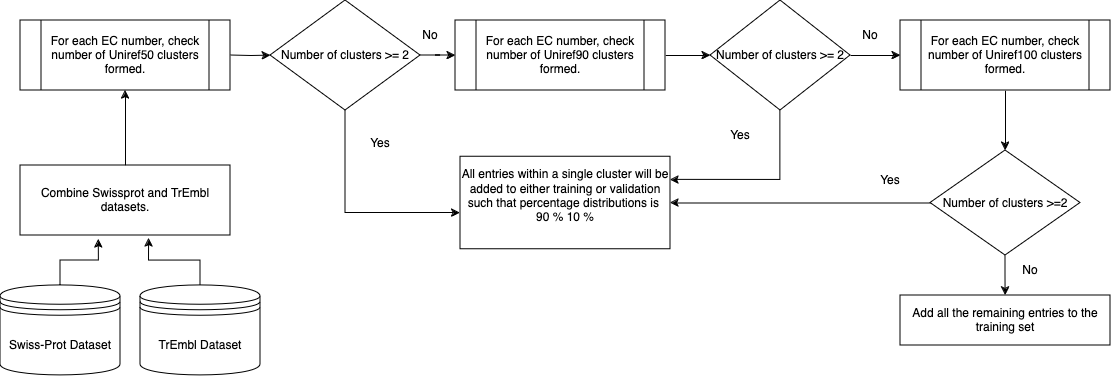}\\
    

            \vspace{3mm}
            \hrule
            \vspace{3mm}
        \flushleft(b) 
                \includegraphics[width=\linewidth]{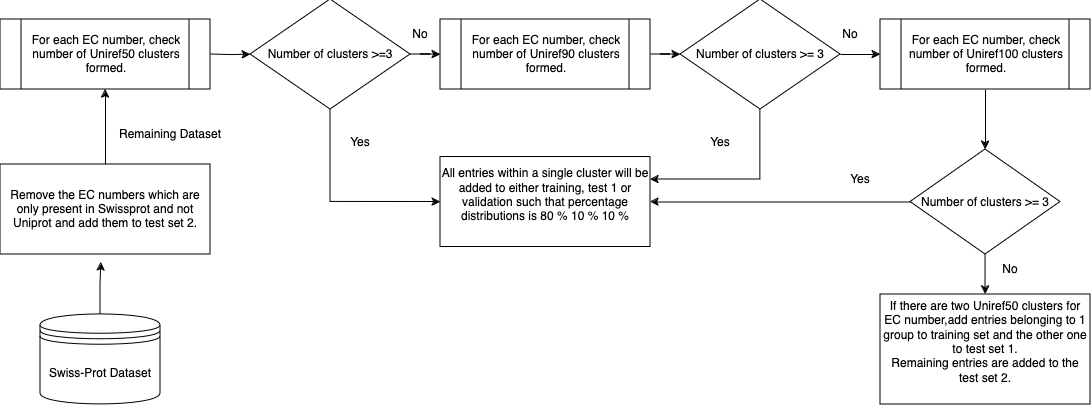}\\
    
                \caption{Splitting pipeline for TrEMBL + SwissProt combined (a) and SwissProt only (b).}
                \label{fig:splitting}
            \end{figure}
            
        \noindent\textbf{Dataset Splitting:} 
            In order to discourage model bias and encourage generalization to novel functions with low sequence similarity to those in biological reference databases, we used a hierarchical strategy to split the dataset into training, validation, and test sets with substantially different sequence identity (Figure \ref{fig:splitting}). This strategy differed slightly for benchmark datasets utilizing SwissProt and TrEMBL combined vs. SwissProt alone (see Section~\ref{subsec:benchmark}) but followed a similar basic splitting procedure: For each unique EC number, we separated sequences into train, valid, and test sets such that no sequence from the same UniRef50 cluster appeared in more than one set. In cases where only one UniRef50 cluster was associated with a specific EC number, we repeated this process based on UniRef90 cluster IDs. If a unique EC number was represented by only one UniRef90 cluster, we then repeated this procedure according to UniRef100 cluster IDs. This hierarchical method allowed us to maintain a high level of sequence diversity within our dataset. By ensuring that the training, validation, and test sets comprised sequences from different clusters, we minimized the risk of overfitting and ensured that the model's performance would generalize well to new, unseen data. 
        
        \noindent\textbf{Natural Language Descriptions:} 
            Our dataset comprises DNA sequences paired with and natural language descriptions providing detailed functional annotations for each entry. Each DNA sequence is associated with an EC number, which categorizes the enzyme by its specific function. The functions performed by the enzymes within a specific EC class are described in natural language  in the KEGG (Kyoto Encyclopedia of Genes and Genomes) database\cite{KEGG, kanehisa2023kegg, kanehisa2019toward}. This information encompasses enzyme classification, hierarchy, the reactions catalyzed by the enzyme, and the general functions of enzymes within that particular class. These annotations collectively offer a more nuanced understanding of enzymatic activities and their broader biological contexts. 
                   
            \begin{figure}[t]
                \centering
                \includegraphics[width=\linewidth]{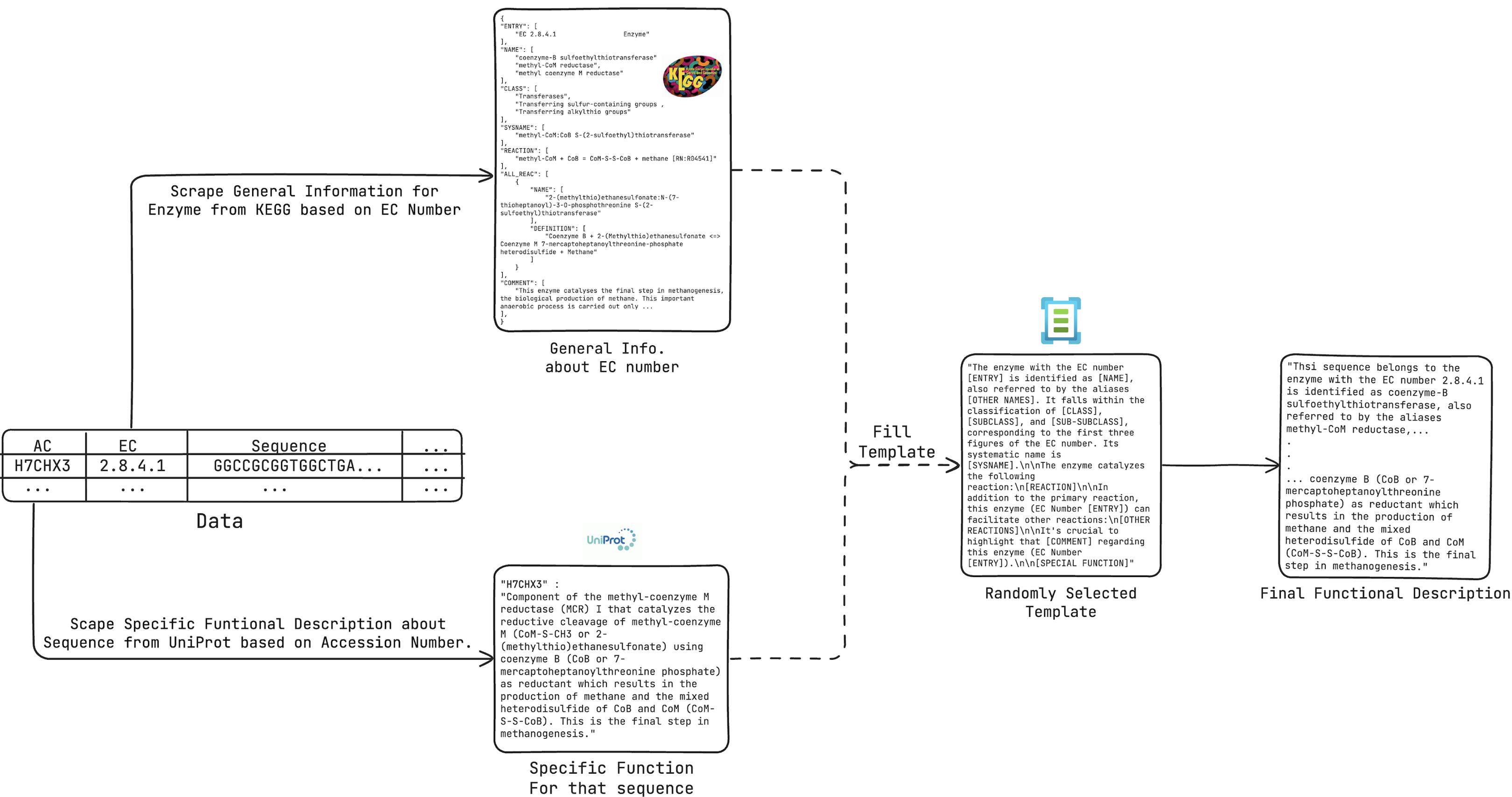}
                \caption{Overview of the description generation pipeline}
                \label{fig:description-flow}
            \end{figure}

            We scraped this information from the Kegg database for each DNA sequence for its corresponding EC number, and further scraped additional functional information from UniProt\cite{uniprot2021} based on their unique protein accession numbers, where available. This included insights into the unique functions that each sequence can perform, providing a more detailed annotation. 
            
            We then converted this information into a natural language functional description for a particular sequence. To ensure the accuracy of the information in each description and avoid potential hallucinations by generative LLMs, we designed ten templates for natural language descriptions with missing values which were filled with the corresponding information scraped from functional annotation databases. A template was randomly selected for each sequence. These templates contained the fields enzyme classification; class, subclass, and sub-subclass of enzymes; catalyzed reactions; functional roles; and specific activities related to the sequences. This structured approach ensures that each DNA sequence in the dataset is accompanied by rich, informative annotations in a textual format that is free from hallucinations.  Figure \ref{fig:description-flow} illustrates our pipeline.
                
    \subsection{Benchmarks}\label{subsec:benchmark} 
        We derived four benchmark datasets. Each dataset contained either only sequences linked to SwissProt, or a union of sequences linked to both TrEMBL and SwissProt; and were either balanced with respect to EC number class label, or left unbalanced as in the original databases. 
        
        \noindent\textbf{Benchmark-I (SwissProt+TrEMBL, unbalanced)}: 
            Gene sequences corresponding to entries from the \textit{TrEMBL} and \textit{Swiss-Prot} datasets were combined to form a single dataset. This combined dataset was then split into training and validation sets using the aforementioned splitting logic. All the records which could not be split were added to the training set. This resulted in approximately 90\% of the data used for training and 10\% for validation.
        
        \noindent\textbf{Benchmark-II (SwissProt+TrEMBL, balanced)}: 
            Leveraging the set created in Benchmark I,
            we employed a systematic approach to balance the dataset based on the counts of entries for each EC number. 
            We start by counting entries for each EC number in the dataset. The median value of these entry counts was then determined. This median value served as the target count for balancing each EC number in the dataset. We applied the balancing logic described below to adjust the number of examples for each EC number to match the target count: for each EC number, if there are more than 250 examples: we evaluated the number of unique clusters formed by the Uniref50 ids. If there are more than 250 clusters, randomly select 250 clusters, and then select one example from each cluster, resulting in exactly 250 examples. If there are fewer than 250 clusters, select one example from each available cluster. The remaining examples needed to reach the count 250 are randomly chosen from the available clusters.
            
            For each EC number, if there are fewer than 250 examples:
            We included all available examples. To reach a total of 250 examples, we upsampled by randomly duplicating examples from the available examples until the target count is achieved.
            By following this method, we ensured that the dataset was balanced according to the median count of entries per EC number, thereby maintaining both diversity and balanced representation of functional classes for robust analysis and comparison.
        
        \noindent\textbf{Benchmark-III (SwissProt only, unbalanced)}:
            This was created exclusively using the Swiss-Prot dataset. 
            We identified 408 EC numbers that are present in Swiss-Prot but absent in UniProt. Entries associated with these EC numbers were removed from the main dataset and set aside to form an \textit{out-of-distribution} test set (Test Set-II).
            
            The remaining Swiss-Prot dataset was split into training, validation, and test sets using the previously described splitting logic. The distribution of data across these sets was Training Set: 80\%, Validation Set: 10\%, Test Set: 10\%. The test set generated from this split was designated as the \textit{in-distribution} test set (Test Set-I). All the records that could not be split were added to the out-of-distribution test set (Test Set-II).
        
        \noindent\textbf{Benchmark-IV (SwissProt only, balanced)}: 
            We utilized the training and validation sets generated from Benchmark III. To ensure balanced representation within these sets, we applied the balancing logic described in Benchmark II, but with a crucial adjustment: the target value for balancing was set to 10. This adjustment allowed us to uniformly balance the data by selecting or duplicating examples accordingly, ensuring each EC number was represented by exactly 10 examples.
    \begin{table}[t]
        \centering
        \caption{\centering{Each benchmark with the number of records and unique enzyme classes.}}
        \begin{adjustbox}{width=\textwidth}
        \begin{tabular}{lcccccccc}
            \toprule
                \multirow{3}{*}{\textbf{Dataset}} & \multicolumn{2}{c}{\textbf{Train}} & \multicolumn{2}{c}{\textbf{Validation}} & \multicolumn{2}{c}{\textbf{Test-1}} & \multicolumn{2}{c}{\textbf{Test-2}} \\
                
                \cmidrule(lr){2-3} \cmidrule(lr){4-5} \cmidrule(lr){6-7} \cmidrule(lr){8-9}
                
                & \textbf{Records} & \textbf{Classes} & \textbf{Records} & \textbf{Classes} & \textbf{Records} & \textbf{Classes}& \textbf{Records} & \textbf{Classes}\\
            \midrule
                Benchmark I & 27,877,140 & 4,684 & 5,370,250 & 4,226 & \multirow{4}{*}{19,930} & \multirow{4}{*}{2,228} & \multirow{4}{*}{4,548} & \multirow{4}{*}{1,196}\\
                Benchmark II & 1,171,000 & 4,684 & 1,056,500 & 4,226 & & & &\\
                Benchmark III & 151,314 & 2,228 & 19,296 & 1,842 & & & & \\
                Benchmark IV & 22,280 & 2,228 & 18,420 & 1,842 & & & &\\
            \bottomrule
        \end{tabular}
        \label{tab:dataset-summary-split}
        \end{adjustbox}
    \end{table}
    
        \noindent\textbf{Test Set-I (In-Distribution Test Set)}:
            The test sets are purely generated from Uniprot/Swiss-Prot dataset.
            The In-Distribution Test Set contains unique sequences with EC functional labels which appear in all four training sets. To ensure that all EC Numbers in this test set appear in training sets of all benchmarks, we removed records with EC numbers from Swiss-Prot which were not present in TrEMBL. 

        \noindent\textbf{Test Set-II (Out-of-Distribution Test Set)}:
            To assess the performance of the models on unseen or rare classes, we generated Test Set-II.
            The Out-of-Distribution Test Set contains EC numbers which do not appear in any of the training sets. This corresponds to EC numbers that appear only in SwissProt, but not in TrEMBL, or for EC Numbers which could not be split by UniRef-based splitting logic in Benchmark III.

    \subsection{Evaluation metrics}\label{sec:metrics}
        To assess the quality of this dataset, we evaluated the four benchmark datasets using standard hierarchical metrics, reflecting the hierarchical nature of the data. This evaluation required high-quality DNA embedding models to generate robust encoder representations. We employed three Transformer-based models for this purpose:
        \textbf{DNABERT}\cite{ji2021dnabert}, \textbf{Nucleotide Transformers}\cite{dalla2023nucleotide} and \textbf{LOLBERT (Language of Life BERT)}\cite{hoarfrost2022deep}.
        


    \noindent\textbf{Hierarchical Precision, Recall and F-Score.}\label{subsec:hp}
    Due to the hierarchical nature of the protein function classes, we use the hierarchical version of precision (hPrecision), recall (hRecall), and F-score (hF-Score) defined in \cite{JMLR:v24:21-1518} to measure classification performance. 

    \noindent\textbf{Top K-NN retrieval}.
    \label{subsec:top-knn}
        To evaluate the clustering quality of DNA embeddings generated by different LLM models, we employed Top K-Nearest Neighbor (K-NN) retrieval\cite{douze2024faiss}. We tested the K-NN retrieval for various values of K (1, 3, and 5) on different embedding types: Class token, Min token, Max token, and Average Token embeddings. The retrieved neighbors for each embedding were then evaluated using the Hierarchical Precision Score (HP) as defined in Section~\ref{subsec:hp}. For each retrieved neighbor, the maximum overlap score was used for the HP calculation. In our particular case, the values of hP, hR, and hF were identical, so we focused solely on HP for analysis.
    
    \noindent\textbf{Cluster Quality Metric}.
    \label{subsec:ClusterQuality}
        To further assess the quality of the embeddings, we employed the silhouette score from the scikit-learn package. This metric helps evaluate how well the embeddings cluster based on the true EC number labels. The silhouette score\cite{shahapure2020cluster} is calculated for each data point by comparing the average distance to other points within its assigned cluster (a) to the average distance to the points in the next nearest cluster (b). The score is then calculated as (b - a) / max(a, b), ranging from -1 to 1. Higher silhouette scores indicate better separation and compactness of clusters.

\section{Baselines and results}
In this section, we train and evaluate deep learning algorithms over the four benchmarks. 
Experiment details including the training procedure and model definitions can be found on the project website. 

\subsection{Top K-NN Retrieval Hierarchical Precision:} 
\label{subsec: Top K-NN Retrieval}
The results of the top-K Nearest Neighbor (k-NN) retrievals for each benchmark dataset are presented in Table~\ref{tab:hierarchical_precision} (refer to Section~\ref{subsec:top-knn} for details on the k-NN methodology). We evaluated the retrievals using Class token embeddings for each benchmark. The results demonstrate that the Finetuned LOLBERT model consistently outperforms all three pre-trained models (LOLBERT, DNABERT, and Nucleotide Transformer) across all benchmarks. Additionally, unbalanced datasets yielded slightly higher hierarchical precision scores compared to balanced datasets.

\begin{table}[t]
    \centering
    \caption{\textbf{Hierarchical Precision of K-NN Retrievals:} This table presents retrieval results
    on class token embeddings generated by four models.
    }
    \label{tab:hierarchical_precision}
    \begin{adjustbox}{width=\textwidth}
    \begin{tabular}{lccccc}
        \toprule
        & \textbf{k} & \textbf{LOLBERT} & \textbf{DNABERT} & \textbf{Nucleotide Transformer}& \textbf{FinetunedLOLBERT} \\
        \midrule
        \multirow{3}{*}{Benchmark I} & 1 & 0.5103 & 0.4143 & 0.4419 & 0.6564\\
        & 3 & 0.5870 & 0.4923 & 0.5480 & 0.7128\\
        & 5 & 0.6258 & 0.5368 & 0.6027 & 0.7382\\
        \midrule
        \multirow{3}{*}{Benchmark II} & 1 & 0.2700 & 0.2215 & 0.2598 & 0.4777\\
        & 3 & 0.3461 & 0.2887 & 0.3424 & 0.5608\\
        & 5 & 0.3879 & 0.3320 & 0.3903 & 0.6000\\
        \midrule
        \multirow{3}{*}{Benchmark III} & 1 & 0.1332 & 0.1168 & 0.1553 & 0.3329\\
        & 3 & 0.2038 & 0.1903 & 0.2380 & 0.4195\\
        & 5 & 0.2416 & 0.2321 & 0.2816 & 0.4635\\
        \midrule
        \multirow{3}{*}{Benchmark IV} & 1 & 0.2746 & 0.2098 & 0.2293 & 0.4751\\
        & 3 & 0.3698 & 0.3087 & 0.2998 & 0.5549\\
        & 5 & 0.4222 & 0.3628 & 0.3439 & 0.5893\\ 
        \bottomrule
    \end{tabular}
    \end{adjustbox}
\end{table}

\subsection{Silhouette Scores across Benchmark:} 
\label{subsec: Silhouette Scores}
The silhouette score is a metric used in cluster analysis to assess the separation and compactness of clusters. Higher scores indicate that points within a cluster are more similar to each other than they are to points in neighboring clusters. Table~\ref{tab:silhouette_score} presents the silhouette scores for the four DNA embedding models evaluated across all four benchmark datasets. Consistent with the k-NN retrieval results, the Finetuned LOLBERT model exhibits the highest silhouette scores across all benchmarks, suggesting that it effectively clusters the embeddings based on their Enzyme Commission (EC) numbers.

\begin{table}[ht]
    \centering
    \caption{\textbf{Silhouette Scores of class embeddings:} 
    Higher scores indicate better clustering quality.}
    \label{tab:silhouette_score}
    \begin{adjustbox}{width=\textwidth}
    \begin{tabular}{lcccc}
        \toprule
         & \textbf{LOLBERT} & \textbf{DNABERT} & \textbf{Nucleotide Transformer}& \textbf{FinetunedLOLBERT} \\
        \midrule
        Benchmark I & -0.2139 & -0.3078 & -0.2237 & -0.1735 \\
        Benchmark II & -0.0672 & -0.1401 & -0.0702 & -0.0466 \\
        Benchmark III & 0.2623 & 0.2009 & 0.2461 & 0.3060 \\
        Benchmark IV & -0.1855 & -0.2900 & -0.2248 & -0.0786 \\
        \bottomrule
    \end{tabular}
    \end{adjustbox}
\end{table}

\subsection{EC Number Prediction using DNA Embedding across Benchmark Datasets:}

Considering the unsupervised embedding results from Section~\ref{subsec: Top K-NN Retrieval} and Section~\ref{subsec: Silhouette Scores}, we selected the finetuned LOLBERT model as the embedding model and trained a two-layer classifier to predict the EC numbers in test sets I and II. After the cross-validation process on the validation dataset, we chose a batch size of 64, a hidden size of 256, a learning rate of 0.001, and trained the model for 10 epochs. The metrics of prediction are presented in Table~\ref{tab: EC_unimodal_predict}. We evaluated the results using hierarchical Precision, Recall (hRecall), and F-Score (hF-Score). The results demonstrate that the unbalanced datasets yielded slightly higher prediction performance. The datasets, including UniProtKB/TrEMBL and UniProtKB/SwissProt, have larger sample sizes and yield better prediction performance. Additionally, for all models trained based on the four benchmark datasets, the prediction ability for the out-of-distribution dataset is not as robust as for the in-distribution test set.

\begin{table}[ht]
\centering
\caption{\textbf{EC Numbers Classification Using Finetuned LOLBERT Model:} 
We train a two-layer MLP from Finetuned LOLBERT embeddings to predict EC numbers.
}
    \label{tab: EC_unimodal_predict}
    \begin{tabular}{lcccc}
    \toprule
    &             & \textbf{hPrecision} & \textbf{hRecall} & \textbf{hF-Score}\\
    \midrule
    \multirow{2}{*}{Benchmark I} & Test-Set-I  & 0.7512 & 0.1878   & 0.3004 \\
                                 & Test-Set-II & 0.4737 & 0.1184   & 0.1894 \\
    \midrule
    \multirow{2}{*}{Benchmark II} & Test-Set-I & 0.5389 & 0.1347   & 0.2155 \\
                                  & Test-Set-II& 0.3960 & 0.0990   & 0.1584 \\
    \midrule
    \multirow{2}{*}{Benchmark III} & Test-Set-I & 0.3097& 0.0774   & 0.1239 \\
                                   & Test-Set-II& 0.2285& 0.0571   & 0.0914 \\
    \midrule
    \multirow{2}{*}{Benchmark IV}  & Test-Set-I & 0.2248& 0.0562   & 0.0899 \\
                                   & Test-Set-II& 0.2146& 0.0536   & 0.0858 \\
    \bottomrule    
\end{tabular}
\end{table}

\subsection {Multi-modal Zero- and Few-shot EC Number Predictions Using LLM Prompts:}

\begin{table}[ht]
\centering
\caption{\textbf{EC Numbers Classification Using Multi-Modal Data:} 
We use Llama3 \cite{llama3modelcard} with both DNA sequences and text descriptions and test its classification performance.
}
\label{tab:llm_predict}
\begin{tabular}{cccccc}
  \toprule
                             &           & \textbf{hPrecision} & \textbf{hRecall}   & \textbf{hF-Score}\\ 
                             \midrule
\multirow{2}{*}{Test-Set-I}  & Zero-Shot & 0.7429. & 0.1689 & 0.3016 \\  
                             & Few-Shot  & 0.8508 & 0.1702 & 0.3243 \\ 
                             \midrule
\multirow{2}{*}{Test-Set-II} & Zero-Shot & 0.7264 & 0.1579 & 0.2976   \\ 
                             & Few-Shot  & 0.8338   & 0.1634   & 0.3135  \\ 
                             \bottomrule
\end{tabular}
\end{table}
To leverage the multi-modal properties of our benchmark datasets, we utilized both DNA sequences and text descriptions to predict EC numbers, employing the open-access Llama 3 language model\cite{touvron2023llama}. For zero-shot prompting, we provided natural language instructions that describe the task and specify the expected output, allowing the LLMs to construct a context that refines the inference space for more accurate outputs. In contrast, few-shot learning, as demonstrated by numerous studies \cite{brown2020language, ahuja-etal-2023-mega}, offers superior performance. In our few-shot experiments with Llama 3, we selected examples from the available training data, simplified the instructions, and included three-shot examples. The results, presented in Table~\ref{tab:llm_predict}, indicate that using multi-modal data improves prediction performance compared to using only DNA embeddings, and that few-shot prompting outperforms zero-shot prompting due to the inclusion of examples. However, more accurate methods for dealing with multi-modal datasets, including DNA sequences and text language, are still needed.

\section{Conclusions}\label{sec:conclusion}
In this work, we propose a novel multi-modal dataset and benchmark suite for enzymatic function prediction and description of DNA sequences. Our data enables the joint learning of functional information encoded in biological sequences themselves alongside the scientific knowledge of functional behavior and mechanism captured in natural language in biological databases. Multiple benchmark datasets are designed and evaluated on both unsupervised and fully supervised function classification tasks. We demonstrate baseline performance with state-of-the-art DNA embedding models as well as large foundation models. Notably, the Finetuned LOLBERT model exhibited superior performance, outperforming three other pre-trained models—LOLBERT, DNABERT, and the Nucleotide Transformer—in all unsupervised tasks. This is an encouraging result indicating that additional effort training multi-modal models on the specific task of functional annotation will yield still better performance. 

\noindent\textbf{Future work}. To enhance our benchmark suite, we plan to incorporate additional tasks and models. Future expansions will include generative tasks, such as training models capable of producing textual descriptions from provided DNA sequences. Moreover, we will explore more sophisticated baseline models, specifically multi-modal deep learning models using contrastive loss to learn embeddings from two modalities simultaneously \cite{chen2020simple}. We are committed to advancing the understanding of enzymatic functions and seek collaboration with the research community to further this field.


\noindent\textbf{Broader impact}. This work establishes a comprehensive multi-modal benchmark that integrates DNA sequences with natural language to enhance enzymatic function understanding. Through a series of unsupervised and supervised tasks, this benchmark assesses the effectiveness of various DNA sequence encoding methods and state-of-the-art large language models in processing biological multi-modal inputs. By evaluating these approaches, we aim to determine their capability to synergistically handle DNA and linguistic data, and gauge their potential applicability in real-world scenarios. This benchmark will provide a robust foundation for applying multi-modal deep learning techniques to biological systems.


{
\bibliographystyle{ieee_fullname}
\bibliography{egbib}
}

\newpage
\section*{Supplementary Materials}
\subsection*{Datasheet for dataset}

\noindent
\textbf{
This document is based on \textit{Datasheets for Datasets} by Gebru \textit{et al.} \cite{gebru2021datasheets}. Please see the most updated version
\underline{\textcolor{blue}{\href{http://arxiv.org/abs/1803.09010}{here}}}.
} \\

\subsection*{MOTIVATION}

    \textcolor{\sectioncolor}{\textbf{
    For what purpose was the dataset created?
    }
    Was there a specific task in mind? Was there
    a specific gap that needed to be filled? Please provide a description.
    } \\
    Predicting gene function from its DNA sequence is a fundamental challenge in biology. Many deep learning models have been proposed to embed DNA sequences and predict their enzymatic function, leveraging information in public databases linking DNA sequences to an enzymatic function label. However, much of the scientific community's knowledge of biological function is not represented in these categorical labels, and is instead captured in unstructured text descriptions of mechanisms, reactions, and enzyme behavior. Deep learning of models predicting enzymatic function are likely to benefit from incorporating this multi-modal data encoding scientific knowledge of biological function. There is, however,  no dataset designed for machine learning algorithms to leverage this multi-modal information. Here we propose a novel dataset and benchmark suite that enables the exploration and development of large multi-modal neural network models on gene DNA sequences and natural language descriptions of gene function.
    \\
    
    \textcolor{\sectioncolor}{\textbf{
    Who created this dataset (e.g., which team, research group) and on behalf
    of which entity (e.g., company, institution, organization)?
    }
     \\
    We are a team of marine science and computer science researchers from the University of Georgia. \\
  }  
    \textcolor{\sectioncolor}{\textbf{
    What support was needed to make this dataset?
    }
    (e.g.who funded the creation of the dataset? If there is an associated
    grant, provide the name of the grantor and the grant name and number, or if
    it was supported by a company or government agency, give those details.)
    } \\
    The creation of this dataset was supported by the University of Georgia.
    
    \textcolor{\sectioncolor}{\textbf{
    Any other comments?
    }} \\
    None. \\

\subsection*{COMPOSITION}

    \textcolor{\sectioncolor}{\textbf{
    What do the instances that comprise the dataset represent (e.g., documents,
    photos, people, countries)?
    }
    Are there multiple types of instances (e.g., movies, users, and ratings;
    people and interactions between them; nodes and edges)? Please provide a
    description.
    } \\
    Instances in the data represent DNA sequence fragments representing shotgun sequencing read-length subsets of genes, function classification labels (in the form of Enzyme Commission numbers), as well as natural language descriptions of enzymes.   \\
    
    \textcolor{\sectioncolor}{\textbf{
    How many instances are there in total (of each type, if appropriate)?
    }
    } \\
    In the training set, Benchmark I contains 27,877,140 DNA sequence records and an equal number of descriptions, representing 4,684 EC classes; Benchmark II contains 1,171,000 DNA sequence records and descriptions, representing 4,684 EC classes; Benchmark III contains 151,314 DNA sequences and descriptions representing 2,228 EC classes and Benchmark IV contains 22,280 DNA sequences and descriptions representing 2,228 EC classes. In the validation set, Benchmark I contains 5,370,250 sequences and descriptions representing 4,226 EC classes; Benchmark II contains 1,056,500 sequences and descriptions representing 4,226 EC classes; Benchmark III contains 19,296 sequences and descriptions representing 1,842 EC classes; and Benchmark IV contains 18,420 sequences and descriptions representing 1,842 EC classes. Test-1 contains 19,930 sequences and descriptions representing 2,228 EC classes; and Test-2 contains 4,548 sequences and descriptions representing 1,196 EC classes. 
    
    \textcolor{\sectioncolor}{\textbf{
    Does the dataset contain all possible instances or is it a sample (not
    necessarily random) of instances from a larger set?
    }
    If the dataset is a sample, then what is the larger set? Is the sample
    representative of the larger set (e.g., geographic coverage)? If so, please
    describe how this representativeness was validated/verified. If it is not
    representative of the larger set, please describe why not (e.g., to cover a
    more diverse range of instances, because instances were withheld or
    unavailable).
    } \\
    This dataset contains DNA sequences and their EC numbers from the UniProt and SwissProt databases. It is an enhanced subset of UniProt and SwissProt.
     \\
    
    \textcolor{\sectioncolor}{\textbf{
    What data does each instance consist of?
    }
    “Raw” data (e.g., unprocessed text or images) or features? In either case,
    please provide a description.
    } \\
    DNA sequence, EC number, language description
     \\
    
    \textcolor{\sectioncolor}{\textbf{
    Is there a label or target associated with each instance?
    }
    If so, please provide a description.
    } \\
    Yes, we use the EC number and language description as the label for the DNA sequence. \\
    
    \textcolor{\sectioncolor}{\textbf{
    Is any information missing from individual instances?
    }
    If so, please provide a description, explaining why this information is
    missing (e.g., because it was unavailable). This does not include
    intentionally removed information, but might include, e.g., redacted text.
    } \\
    None. \\
    
    \textcolor{\sectioncolor}{\textbf{
    Are relationships between individual instances made explicit (e.g., users’
    movie ratings, social network links)?
    }
    If so, please describe how these relationships are made explicit.
    } \\
    Each instance is a DNA sequence. Relations between individual instances are reflected in their EC number. \\
    
    \textcolor{\sectioncolor}{\textbf{
    Are there recommended data splits (e.g., training, development/validation,
    testing)?
    }
    If so, please provide a description of these splits, explaining the
    rationale behind them.
    } \\
    Yes, the training and testing splits for each of the benchmarks are described in the main document. \\
    
    \textcolor{\sectioncolor}{\textbf{
    Are there any errors, sources of noise, or redundancies in the dataset?
    }
    If so, please provide a description.
    } \\
    None. \\
    
    \textcolor{\sectioncolor}{\textbf{
    Is the dataset self-contained, or does it link to or otherwise rely on
    external resources (e.g., websites, tweets, other datasets)?
    }
    If it links to or relies on external resources, a) are there guarantees
    that they will exist, and remain constant, over time; b) are there official
    archival versions of the complete dataset (i.e., including the external
    resources as they existed at the time the dataset was created); c) are
    there any restrictions (e.g., licenses, fees) associated with any of the
    external resources that might apply to a future user? Please provide
    descriptions of all external resources and any restrictions associated with
    them, as well as links or other access points, as appropriate.
    } \\
    The dataset is self-contained. A model can be trained fully with the data we provided. \\
    
    \textcolor{\sectioncolor}{\textbf{
    Does the dataset contain data that might be considered confidential (e.g.,
    data that is protected by legal privilege or by doctor-patient
    confidentiality, data that includes the content of individuals’ non-public
    communications)?
    }
    If so, please provide a description.
    } \\
    No. \\
    
    \textcolor{\sectioncolor}{\textbf{
    Does the dataset contain data that, if viewed directly, might be offensive,
    insulting, threatening, or might otherwise cause anxiety?
    }
    If so, please describe why.
    } \\
    No. \\
    
    \textcolor{\sectioncolor}{\textbf{
    Does the dataset relate to people?
    }
    If not, you may skip the remaining questions in this section.
    } \\
    No. \\
    
    \textcolor{\sectioncolor}{\textbf{
    Does the dataset identify any subpopulations (e.g., by age, gender)?
    }
    If so, please describe how these subpopulations are identified and
    describe their respective distributions within the dataset.
    } \\
    N/A. \\
    
    \textcolor{\sectioncolor}{\textbf{
    Is it possible to identify individuals (i.e., one or more natural persons),
    either directly or indirectly (i.e., in combination with other data) from
    the dataset?
    }
    If so, please describe how.
    } \\
    N/A. \\
    
    \textcolor{\sectioncolor}{\textbf{
    Does the dataset contain data that might be considered sensitive in any way
    (e.g., data that reveals racial or ethnic origins, sexual orientations,
    religious beliefs, political opinions or union memberships, or locations;
    financial or health data; biometric or genetic data; forms of government
    identification, such as social security numbers; criminal history)?
    }
    If so, please provide a description.
    } \\
    N/A. \\
    
    \textcolor{\sectioncolor}{\textbf{
    Any other comments?
    }} \\
    None. \\

 \subsection*{COLLECTION}
    \textcolor{\sectioncolor}{\textbf{
    How was the data associated with each instance acquired?
    }
    Was the data directly observable (e.g., raw text, movie ratings),
    reported by subjects (e.g., survey responses), or indirectly
    inferred/derived from other data (e.g., part-of-speech tags, model-based
    guesses for age or language)? If data was reported by subjects or
    indirectly inferred/derived from other data, was the data
    validated/verified? If so, please describe how.
    } \\
    We use the DNA sequences from the UniProt database. \\
    
    \textcolor{\sectioncolor}{\textbf{
    Over what timeframe was the data collected?
    }
    Does this timeframe match the creation timeframe of the data associated
    with the instances (e.g., recent crawl of old news articles)? If not,
    please describe the timeframe in which the data associated with the
    instances was created. Finally, list when the dataset was first published.
    } \\
    The data in UniProt is continuously collected by the UniProt consortium since 2002. We accessed data records from the 2021 version of the database.  \\
    
    \textcolor{\sectioncolor}{\textbf{
    What mechanisms or procedures were used to collect the data (e.g., hardware
    apparatus or sensor, manual human curation, software program, software
    API)?
    }
    How were these mechanisms or procedures validated?
    } \\
    We downloaded the data from the official UniProt database via their APIs. \\
    
    \textcolor{\sectioncolor}{\textbf{
    What was the resource cost of collecting the data?
    }
    (e.g. what were the required computational resources, and the associated
    financial costs, and energy consumption - estimate the carbon footprint.
    See Strubell \textit{et al.}\cite{strubellEnergyPolicyConsiderations2019} for approaches in this area.)
    } \\
    Data were downloaded via the UniProt API and processed into benchmark datasets on a compute node equipped with an 80GB A100 GPU node. Processing took approximately 24 node hours, corresponding  to 4.15 kg CO2 eq. 
    
    \textcolor{\sectioncolor}{\textbf{
    If the dataset is a sample from a larger set, what was the sampling
    strategy (e.g., deterministic, probabilistic with specific sampling
    probabilities)?
    }
    } \\
    The dataset is a subset of UniProt. The data filtering and sampling strategy is described in the main document Section 3.1. \\
    
    \textcolor{\sectioncolor}{\textbf{
    Who was involved in the data collection process (e.g., students,
    crowdworkers, contractors) and how were they compensated (e.g., how much
    were crowdworkers paid)?
    }
    } \\
    Grad students in the School of Computing at the University of Georgia were involved in the data collection process. \\
    
    \textcolor{\sectioncolor}{\textbf{
    Were any ethical review processes conducted (e.g., by an institutional
    review board)?
    }
    If so, please provide a description of these review processes, including
    the outcomes, as well as a link or other access point to any supporting
    documentation.
    } \\
    N/A. \\
    
    \textcolor{\sectioncolor}{\textbf{
    Does the dataset relate to people?
    }
    If not, you may skip the remainder of the questions in this section.
    } \\
    No. \\

    \textcolor{\sectioncolor}{\textbf{
    Any other comments?
    }} \\
    None. \\

\subsection*{PREPROCESSING / CLEANING / LABELING}

    \textcolor{\sectioncolor}{\textbf{
    Was any preprocessing/cleaning/labeling of the data
    done(e.g.,discretization or bucketing, tokenization, part-of-speech
    tagging, SIFT feature extraction, removal of instances, processing of
    missing values)?
    }
    If so, please provide a description. If not, you may skip the remainder of
    the questions in this section.
    } \\
    Filtering and cleaning of the data has been conducted. The detailed process is in the main document Section 3.1.  \\

    \textcolor{\sectioncolor}{\textbf{
    Was the “raw” data saved in addition to the preprocessed/cleaned/labeled
    data (e.g., to support unanticipated future uses)?
    }
    If so, please provide a link or other access point to the “raw” data.
    } \\
    The raw data is stored in the internal storage by our research lab. \\

    \textcolor{\sectioncolor}{\textbf{
    Is the software used to preprocess/clean/label the instances available?
    }
    If so, please provide a link or other access point.
    } \\
    No special software was used. \\

    \textcolor{\sectioncolor}{\textbf{
    Any other comments?
    }} \\
    None. \\

\subsection*{USES}
    \textcolor{\sectioncolor}{\textbf{
    Has the dataset been used for any tasks already?
    }
    If so, please provide a description.
    } \\
    No. \\

    \textcolor{\sectioncolor}{\textbf{
    Is there a repository that links to any or all papers or systems that use the dataset?
    }
    If so, please provide a link or other access point.
    } \\
    \url{https://github.com/hoarfrost-lab/biotalk} \\

    \textcolor{\sectioncolor}{\textbf{
    What (other) tasks could the dataset be used for?
    }
    } \\
    The dataset could be used for example to generate DNA sequence from text descriptions.\\

    \textcolor{\sectioncolor}{\textbf{
    Is there anything about the composition of the dataset or the way it was
    collected and preprocessed/cleaned/labeled that might impact future uses?
    }
    For example, is there anything that a future user might need to know to
    avoid uses that could result in unfair treatment of individuals or groups
    (e.g., stereotyping, quality of service issues) or other undesirable harms
    (e.g., financial harms, legal risks) If so, please provide a description.
    Is there anything a future user could do to mitigate these undesirable
    harms?
    } \\
    None. \\

    \textcolor{\sectioncolor}{\textbf{
    Are there tasks for which the dataset should not be used?
    }
    If so, please provide a description.
    } \\
    None. \\

    \textcolor{\sectioncolor}{\textbf{
    Any other comments?
    }} \\
    None. \\

\subsection*{DISTRIBUTION}
    \textcolor{\sectioncolor}{\textbf{
    Will the dataset be distributed to third parties outside of the entity
    (e.g., company, institution, organization) on behalf of which the dataset
    was created?
    }
    If so, please provide a description.
    } \\
    The dataset will be publicly available. \\

    \textcolor{\sectioncolor}{\textbf{
    How will the dataset will be distributed (e.g., tarball on website, API,
    GitHub)?
    }
    Does the dataset have a digital object identifier (DOI)?
    } \\
    The dataset will be downloadable from Google Drive and the code can be obtained on GitHub. \\

    \textcolor{\sectioncolor}{\textbf{
    When will the dataset be distributed?
    }
    } \\
    The dataset is available on the project website: \url{https://github.com/hoarfrost-lab/biotalk}. \\

    \textcolor{\sectioncolor}{\textbf{
    Will the dataset be distributed under a copyright or other intellectual
    property (IP) license, and/or under applicable terms of use (ToU)?
    }
    If so, please describe this license and/or ToU, and provide a link or other
    access point to, or otherwise reproduce, any relevant licensing terms or
    ToU, as well as any fees associated with these restrictions.
    } \\
    The dataset and code are under the Creative Commons Attribution 4.0 License. \\

    \textcolor{\sectioncolor}{\textbf{
    Have any third parties imposed IP-based or other restrictions on the data
    associated with the instances?
    }
    If so, please describe these restrictions, and provide a link or other
    access point to, or otherwise reproduce, any relevant licensing terms, as
    well as any fees associated with these restrictions.
    } \\
    None. \\

    \textcolor{\sectioncolor}{\textbf{
    Do any export controls or other regulatory restrictions apply to the
    dataset or to individual instances?
    }
    If so, please describe these restrictions, and provide a link or other
    access point to, or otherwise reproduce, any supporting documentation.
    } \\
    None.\\

    \textcolor{\sectioncolor}{\textbf{
    Any other comments?
    }} \\
    None. \\

\subsection*{MAINTENANCE}

    \textcolor{\sectioncolor}{\textbf{
    Who is supporting/hosting/maintaining the dataset?
    }
    } \\
    Profs Adrienne Hoarfrost and Jin Sun will be supporting and maintaining the dataset. \\

    \textcolor{\sectioncolor}{\textbf{
    How can the owner/curator/manager of the dataset be contacted (e.g., email
    address)?
    }
    } \\
    Adrienne Hoarfrost (adrienne.hoarfrost@uga.edu)\\ Jin Sun (jinsun@uga.edu) \\

    \textcolor{\sectioncolor}{\textbf{
    Is there an erratum?
    }
    If so, please provide a link or other access point.
    } \\
    None. \\

    \textcolor{\sectioncolor}{\textbf{
    Will the dataset be updated (e.g., to correct labeling errors, add new
    instances, delete instances)?
    }
    If so, please describe how often, by whom, and how updates will be
    communicated to users (e.g., mailing list, GitHub)?
    } \\
    Any updates of the dataset will be announced on the project's website. \\

    \textcolor{\sectioncolor}{\textbf{
    If the dataset relates to people, are there applicable limits on the
    retention of the data associated with the instances (e.g., were individuals
    in question told that their data would be retained for a fixed period of
    time and then deleted)?
    }
    If so, please describe these limits and explain how they will be enforced.
    } \\
    N/A.\\

    \textcolor{\sectioncolor}{\textbf{
    Will older versions of the dataset continue to be
    supported/hosted/maintained?
    }
    If so, please describe how. If not, please describe how its obsolescence
    will be communicated to users.
    } \\
    Yes. An updated version of the dataset will be kept on the website. \\

    \textcolor{\sectioncolor}{\textbf{
    If others want to extend/augment/build on/contribute to the dataset, is
    there a mechanism for them to do so?
    }
    If so, please provide a description. Will these contributions be
    validated/verified? If so, please describe how. If not, why not? Is there a
    process for communicating/distributing these contributions to other users?
    If so, please provide a description.
    } \\
    Others can extend and build on the dataset under the CC 4.0 license. \\

    \textcolor{\sectioncolor}{\textbf{
    Any other comments?
    }} \\
    None. \\

\section*{Website}

Project website: \url{https://hoarfrost-lab.github.io/BioTalk/}.

GitHub: \url{https://hoarfrost-lab.github.io/BioTalk/}.

\section*{Croissant metadata record}

\url{https://github.com/Hoarfrost-Lab/BioTalk/blob/main/croissant.json}.

\section*{Author statement}
The authors bear all responsibility in case of violation of rights.

\section*{Hosting, licensing, and maintenance plan}
The downloadable link to the dataset is on the project's website. It is a permanent link to a cloud storage space.

The dataset and code are released under a Creative Commons Attribution 4.0 International license: \url{https://creativecommons.org/licenses/by/4.0/legalcode.txt}.

The dataset will be maintained by faculty and graduate students in the Department of Marine Sciences and the School of Computing at the University of Georgia. 
The data storage account is owned by a University of Georgia-sponsored account. The dataset might be updated with new data and/or new labels. Such updates will be announced on the project's website.

\newpage
\section*{Checklist}


\begin{enumerate}

\item For all authors...
\begin{enumerate}
  \item Do the main claims made in the abstract and introduction accurately reflect the paper's contributions and scope?
    \answerYes{Main claims are consistent across different parts of the paper.}
  \item Did you describe the limitations of your work?
    \answerYes{Limitations and future directions are discussed in Section \ref{sec:conclusion}.}
  \item Did you discuss any potential negative societal impacts of your work?
    \answerNA{}
  \item Have you read the ethics review guidelines and ensured that your paper conforms to them?
    \answerYes{}
\end{enumerate}

\item If you are including theoretical results...
\begin{enumerate}
  \item Did you state the full set of assumptions of all theoretical results?
    \answerNA{}
	\item Did you include complete proofs of all theoretical results?
    \answerNA{}
\end{enumerate}

\item If you ran experiments (e.g. for benchmarks)...
\begin{enumerate}
  \item Did you include the code, data, and instructions needed to reproduce the main experimental results (either in the supplemental material or as a URL)?
    \answerYes{All code and data are in the project website.}
  \item Did you specify all the training details (e.g., data splits, hyperparameters, how they were chosen)?
    \answerYes{}
	\item Did you report error bars (e.g., with respect to the random seed after running experiments multiple times)?
    \answerNA{We use standard metrics in classification tasks.}
	\item Did you include the total amount of compute and the type of resources used (e.g., type of GPUs, internal cluster, or cloud provider)?
    \answerYes{}
\end{enumerate}

\item If you are using existing assets (e.g., code, data, models) or curating/releasing new assets...
\begin{enumerate}
  \item If your work uses existing assets, did you cite the creators?
    \answerYes{}
  \item Did you mention the license of the assets?
    \answerYes{License information is on the project website.}
  \item Did you include any new assets either in the supplemental material or as a URL?
    \answerYes{}
  \item Did you discuss whether and how consent was obtained from people whose data you're using/curating?
    \answerNA{}
  \item Did you discuss whether the data you are using/curating contains personally identifiable information or offensive content?
    \answerNA{}
\end{enumerate}

\item If you used crowdsourcing or conducted research with human subjects...
\begin{enumerate}
  \item Did you include the full text of instructions given to participants and screenshots, if applicable?
    \answerNA{}
  \item Did you describe any potential participant risks, with links to Institutional Review Board (IRB) approvals, if applicable?
    \answerNA{}
  \item Did you include the estimated hourly wage paid to participants and the total amount spent on participant compensation?
    \answerNA{}
\end{enumerate}

\end{enumerate}



\end{document}